\documentclass{article}
\usepackage{tfrupee} 
\usepackage{textcomp}
\usepackage{amsmath}
\usepackage{amsfonts}
\usepackage{amssymb}
\usepackage{graphicx}
\usepackage{hyperref}
\usepackage{enumitem} 
\usepackage{tfrupee} 
\newlist{PS}{enumerate}{1}
\setlist[PS]{label=\textbf{PS-\arabic*:}}

\newcommand{\footremember}[1]{
	\footnote{#1} 
	\newcounter{#1}
	\setcounter{#1}{\value{footnote}}%
}
 
\title{The Cross-section of Expected Returns on Penny Stocks\\ \textit{Are Low-hanging Fruits Not-so Sweet}?}
\author{%
	Ananjan Bhattacharyya\footremember{Undergraduate Student, Indian Institute of Technology Kharagpur. Email: ananjan.bhattacharya@gmail.com.}
	\and Abhijeet Chandra\footremember{Vinod Gupta School of Management, Indian Institute of Technology Kharagpur. Email: abhijeet@vgsom.iitkgp.ernet.in}
}
\date{}

\begin{document}

\maketitle

\begin{abstract}
	In this paper, we study the determinants of expected returns on the listed penny stocks from two perspectives. Traditionally financial economics literature has been devoted to study the macro and micro determinants of expected returns on stocks ( \href{http://citeseerx.ist.psu.edu/viewdoc/download;jsessionid=D58A0B6276931C46640D9118F6BD21B6?doi=10.1.1.214.7209&rep=rep1&type=pdf}{Subrahmanyam, 2010}). Very few research has been carried out on penny stocks (\href{http://papers.ssrn.com/sol3/papers.cfm?abstract_id=1917300}{Liu, Rhee, \& Zhang, 2011}; \href{http://www.emeraldinsight.com/doi/full/10.1108/RBF-12-04-02}{Nofsinger \& Verma, 2014}). Our study is an effort to contribute more empirical evidence on penny stocks in the emerging market context. We see the return dynamics of penny stocks from corporate governance perspective. Issues such as shareholding patters are considered to be of much significance when it comes to understand the price movements.	Using cross-sectional data on 167 penny stocks listed in the National Stock Exchange of India, we show that (i) Returns of portfolio of lower market-cap penny stocks are significantly different(higher) than that of higher market-cap penny stocks. (ii) Returns of portfolio lower P/E stocks are significantly different (higher) than that of higher P/E stocks. Similarly, returns of portfolio of lower P/B stocks are significantly different (higher) than that of higher P/B stocks, and returns of portfolio of lower priced penny stocks are significantly different(higher) than that of higher priced penny stocks. (iii) Trading volume differences due to alphabetism are insignificant. (iv) Differences in returns of portfolios based on beta and shareholding patterns are insignificant. 
\end{abstract}



\newpage

\section{Introduction}
The literature on financial economics in general and asset pricing in particular has traditionally focused on explaining the prices of financial assets in the context of large-cap, liquid assets which can be traded easily. Originally the single-factor model of asset pricing such as Capital Asset Pricing Model (CAPM) assumes that stocks (or, assets) are easily traded and there is very little price impact against trading. However, over the years multi-factor asset pricing models, such as three-factor and four-factor models, have considered several anomalies such as book-to-market, momentum, market capitalization, liquidity, and size being important ones (see, for example, Banz, 1981; Stattman, 1980; Rosenberg et al., 1985; Fama and French,  1992, 1993, 1995, 1996, 1998, 2000; Jegadeesh and Titman, 1993; Carhart, 1997; Rouwenhorst, 1999). These models have empirically taken into account the risk premium attributed to several factors. The price fluctuations attributed to such risk factors result in the return differentials when compared to stocks of different sizes. The  traditional asset pricing theory suggests that there should be a risk premium for every risk factor and there exists a positive relationship between risk and expected returns. It essentially implies that the higher the risk associated, the higher should be the expected returns on that asset. However, sometimes it is observed that many asset managers generate a negative relationship between risk and return because they raise the volatility of overvalued assets. Socially optimal contracts provide steeper performance incentives and cause larger pricing distortions than privately optimal contracts (\href{http://www.nber.org/papers/w20480}{Buffa et al., 2014}). For the large-cap and mid-cap stocks that are highly liquid in terms of tradability with little price impact, this positive risk-relationship holds true across markets, but it might not remain valid when it comes to stocks that are not so liquid and sought-after by the investors. The stocks which are cheap and less liquid may essentially show an entirely different behavior. With this background, it becomes imperative to understand if this positive risk-return relationship prevails in case of stocks which are cheap, also known as penny stocks.

Usually cheap products and services tempt consumers across markets. In the context of investment and financial decision-making, it is always recommended to \textit{buy low and sell high}. When investors find an opportunity to invest in cheap stocks, they tend to ignore the risk factors associated with the opportunity. Although buying cheap could be expensive (\href{http://www.cs.cornell.edu/~pbriest/papers/udp.pdf}{Briest \& Krysta, 2007}), investors and traders fail to note or choose to ignore fundamentals when investing in cheap stocks. Very cheap stocks, also known as \textit{penny stocks}, are deemed to be illiquid and prone to more risk compared to mid-cap or even small-cap stocks. The pricing of such assets become more complicated in an emerging market such as India, where there might exist information asymmetry and other corporate governance-related issues. Ironically, in the movie "The Wolf of Wall Street", the key protagonist Jordan Belfort, played by Leonardo DeCaprio is shown to have made huge sums of money out of anomalous and fraudulent pricing of penny stocks. While in practice, it's rarely achieved feat for retail investors. In this paper, we attempt to understand if penny stocks are worth investments. Since they are cheap stocks, they're often ignored or discounted off easily. We try to find out if microstructral factors such as trading volume or liquidity, risk characteristics, and market capitalization, and governance-related factors such as company's life (cycle), industry category, shareholding patterns, and so on, influence their expected returns? There have been some instances wherein their prices were easily manipulated by the brokers/traders, and/or the retail investors committed mistakes while trying to understand their risk-return dynamics. We provide empirical evidence in the context of the penny stocks listed in the Indian stock market.

Specifically we study the determinants of expected returns on the listed penny stocks from two perspectives. First, we show the relationship of microstructural factors with expected returns on the penny stocks. Substantial literature in financial economics have been devoted to study the macro and micro determinants of expected returns on stocks (see, for example, \href{http://citeseerx.ist.psu.edu/viewdoc/download;jsessionid=D58A0B6276931C46640D9118F6BD21B6?doi=10.1.1.214.7209&rep=rep1&type=pdf}{Subrahmanyam, 2010}, for a comprehensive review of literature on the issue). Very few research has been carried out on penny stocks (\href{http://papers.ssrn.com/sol3/papers.cfm?abstract_id=1917300}{Liu, Rhee, \& Zhang, 2011}; \href{http://www.emeraldinsight.com/doi/full/10.1108/RBF-12-04-02}{Nofsinger \& Verma, 2014}). Our study is an effort to contribute more empirical evidence on penny stocks in the emerging market context. Secondly, we see the return dynamics of penny stocks from corporate governance perspective. Issues such as shareholding patters are considered to be of much significance when it comes to understand the price movements. This study provides evidence in this context as well.

Using cross-sectional data on 167 penny stocks listed in the National Stock Exchange of India, we employ econometric methodologies to explain the relationship of expected returns on penny stocks with other related variables. We show that (i) Returns of portfolio of lower market-cap penny stocks are significantly different(higher) than that of higher market-cap penny stocks. (ii) Returns of portfolio lower P/E stocks are significantly different(higher) than that of higher P/E stocks. Similarly, returns of portfolio of lower P/B stocks are significantly different (higher) than that of higher P/B stocks, and returns of portfolio of lower priced penny stocks are significantly different (higher) than that of higher priced penny stocks. (iii) Trading volume differences due to alphabetism are insignificant. (iv) Differences in returns of portfolios based on beta and shareholding patterns are insignificant. A series of t-tests is employed to check the statistical relationship between two portfolios (top 50 percentile and bottom 50 percentile) after sorting with respect to the different factors. 

Remaining of the paper is organized as following. Section 2 presents a detailed and comprehensive review of literature relevant to the issues at hand. Precisely we are reviewing recent research carried out in the areas of financial economics, behavioral finance and economics, and asset pricing. In Section 3, detailed description of data variables, measures, and methodological tools used in this study is provided. The basic properties of the data set is also discussed briefly. It also discusses the methods and metrics used for analyzing the data set, along with the research objectives and functional hypotheses. Section 4 provides empirical results and estimations. It also gives discussions and inferences drawn from the analysis. Finally, in Section 5, we present concluding remarks and implications of the study.

\section{Literature Review}

Typically, in asset pricing literature, researchers tend to ignore the small, low-value stocks for examining and explaining the dynamics of risk-return. This may be one of the ways to tackle the scarcity of data related to such stocks. This problem arises mainly because, globally, penny stocks generally trade on the over-the-counter markets, such as the OTC Bulletin Board or the Pink Sheets. However, they may also trade on stock exchanges.  In the United States, a significant number of penny stocks are traded on the electronic securities exchanges, their price movements and
trading activities have important impacts on the whole market. In India, such penny stocks being traded on the stock exchange is very limited. The problem related to data on penny stocks in India becomes even more severe, hence there is hardly any research on Indian penny stocks.

Penny stocks typically suffers with several fundamental and market microstructure-related issues. One of the most significant characteristics of penny stocks is high illiquidity. As mentioned by Liu et al. (2011), usually penny stocks have fewer shareholders; they may not trade as frequently as large cap stocks, and their trading volumes are often very low. This lack of liquidity can cause high price volatility with a sudden change in demand or supply of stocks. The lack of liquidity can
also make it difficult to sell a stock and liquidate the investment returns, especially when there are no buyers that day. Considering the liquidity problem of penny stocks, the Securities \& Exchange Commission (\href{http://www.sec.gov/answers/penny.htm.}{SEC}) warns that, ''Penny stocks may trade infrequently, which means that it may be difficult to
sell penny stock shares once you own them. Because it may be difficult to find quotations
for certain penny stocks, they may be impossible to accurately price. Investors in penny
stocks should be prepared for the possibility that they may lose their whole investment.''

Another important attribute of penny stocks is that they carry huge potential for profits for those investors who are high risk-seekers as these stocks carry high risk mainly because of information asymmetry, low liquidity, and uncertainity related to the fundamentals of such companies. The asymmetric information leads to under-valuation of penny stocks. This essentially enables aggressive investors to find the right kind of penny stocks and presents huge profit potential. While the penny stocks are highly risky and carry high potential for returns, they are cheaply available hence can be invested in easily by even small retail investors who might not have deep pockets. It is, therefore, penny stocks seem a tempting investment choice for small retail investors. However, due to high risk attribute, if wrongly traded, penny stocks might prove detrimental for small retail investors.

Limitations relating to the availability of data on penny stocks have resulted in very few empirical research on such stocks. To quote Nofsinger and Verma (2014), the literature investigating penny stocks or describing the general trading activity of investors who invest in them is sparse. Bollen and Christie (2009) study the market-microstructure of Pink sheet stocks (subset of OTC stocks) and find that market participants have endogenously selected price-dependent tick sizes for different stocks. Hanke and Hauser (2008) study the effects of stock spam e-mails on prices of OTC securities (Pink Sheets and OTC Bulletin Board stocks) and find that positive news contained in stock spam e-mails had no lasting positive effect on stock prices. Harris et al. (2008) find that firms that are delisted from NASDAQ and are relegated to the OTC Bulletin Board and Pink Sheets experience a large decline in liquidity, which is also associated with a significant decline in wealth. Bradley et al. (2006) study penny stock IPOs and find that they have higher initial returns than ordinary IPOs, but significantly worse long-run under-performance. Due to a lack of data, their study did not include IPOs that initially started trading on the OTCBB, Pink Sheets, or the gray market. They do include offerings that initially started trading on the Nasdaq Small Cap Market with an offer price of less than or equal to \$5. Beatty and Kadiyala (2003) find that the Penny Stock Reform Act of 1990 (PSRA) had the cosmetic effect of reducing the number of IPOs priced below \$5, but had no substantive impact on issuer quality.

Penny stocks are attractive investment avenues for small retail investors and most of them are aware that such investments might be wiped out due to huge risk associated with the penny stocks. There exists information idiosyncraticity leading to the likelihood of some investors making abnormal profits while many others losing out their investments.  Kumar (2009) finds that the individual investor’s demand for lottery type stocks (stocks with high idiosyncratic variance, high idiosyncratic skewness, and low prices) increases when economic times are poor and these demand shifts influence the returns of lottery type stocks. Also, Kumar (2009) documents that socio-economic characteristics, like being younger and less affluent, are common for lottery buyers and investors seeking lottery type stocks. 

Several studies have been carried out with the data obtained from the over-the-counter markets. Across the globe, majority of trading in penny stocks happens in the OTC market. For example, Shefrin and Statman (2000) show that behavioral portfolio investors do not mean-variance optimize their entire portfolios. Instead they form separate portfolio layers or mental accounts, each associated with unique goals or aspirations. Behavioral portfolio theory is consistent with an investor holding safer assets for downside protection, while also preferring risky assets to achieve certain wealth aspirations or to satisfy an innate sensation seeking desire. Dorn and Sengmeuller (2009) carried out a detailed survey of German retail investors in order to find out how much they enjoyed investing and gambling. Those who enjoyed them more traded twice as much as those who stated that they did not gamble. Grinblatt and Keloharju (2009) study people in Finland. Their unique data set allow them to merge stock brokerage data with other databases. For example, they know how many speeding tickets their investors have received and have access to psychology tests given during mandatory military service, for males. People who are sensation seeking in one area, like playing poker, tend to also be sensation seeking in other areas. Thus, they can compare the investors’ activities of those people that are prone to sensation seeking (a higher number of speeding tickets) with those who are not. After controlling for other investors characteristics, they find that sensation seekers trade more than other investors. They seem to derive some entertainment from trading. It may be that trading in OTC stocks is simply more fun because of their volatility. Unlike the typical lottery buyers/gamblers, sensation seekers trade OTC stocks primarily for thrill and are not driven as much by economic incentives. Most stocks traded on OTC markets simplylack good public information compared to stocks listed on major stock exchanges. Another possible, though unlikely, explanation for trading OTC stocks is that individuals possess some private or superior information about them.

Because of the unavailability of data pertaining to the trading of penny stocks in the OTC market, we restrict our focus on those penny stocks being traded in the stock exchange. Effectively, we study the risk-return dynamics of penny stocks traded on the National Stock Exchange (NSE) of India. Although the NSE advises its small and retail investors against getting influenced into buying into fundamentally unsound companies (penny stocks\footnote{Penny stocks are those stocks that trade at a very low price, have very low market capitalization, are mostly illiquid, and are usually listed on a smaller exchange. Penny stocks in the Indian stock market can have prices below Rs 10. These stocks are very speculative in nature and are considered highly risky because of lack of liquidity, smaller number of shareholders, large bid-ask spreads and limited disclosure of information. (\href{http://economictimes.indiatimes.com/definition/penny-stock}{The Economic Times})}) based on sudden spurts in trading volumes or prices or favourable articles/stories in the media, it has been seeing growing interest of traders in penny stocks (\href{https://www.nseindia.com/invest/resources/download/prs_guide_english.pdf}{NSE's Investors' Guide to the Capital Market, 2010}).

Another issue that we are studying in this study is what we call as the prevalence of alphabetism in stock market. \href{http://rof.oxfordjournals.org/content/early/2015/04/27/rof.rfv012.full.pdf+html}{Itzkowitz, Itzkowitz, and Rothbort, (\textit{RFS}, 2015)}  show that because stock information is most frequently presented in alphabetical order, stocks with names appearing early in the alphabet (“early alphabet stocks”) will be traded more frequently than stocks with names that begin with letters that appear later in the alphabet (“later alphabet stocks”). In psychology literature, it is shown that that individuals are quasi-rational economic agents trying to satisfy their economic goals in their best possible ways, with some oblivious mistakes that they are most likely to be unaware of. An investor, when faced with a large number of options, often choose the first acceptable option, rather than the best possible option (Simon, 1957). Given the vast quantity of information available and the widespread convention of listing stocks in alphabetical order, it is quite likely that investors are more tempted to buy and sell stocks with early alphabet names. Consistent with this view, Itzkowitz et al. (2015) find that early alphabet stocks are traded more frequently than later alphabet stocks and that alphabeticity also affects firm value. They also document how these effects have changed over time. In this context, we try to examine if such phenomena exists in the small universe of penny stocks listed in the Indian stock exchange. Since the stocks available at dirt cheap prices are very few, we conjecture that investors should not fail to see a complete list of stocks before they undertake trading decisions. Even the casual investor searching an investment website is presented with the possibility of examining thousands of stocks. For each stock, there is substantial information to consider such as the stock’s price, 52-week high and low price, dividend, year-to-date return, and volatility. Ideally, when making an investment, one should investigate all of the information about each possible opportunity before deciding. However, due to limits inherent in humans’ cognitive capacity, and given the vast quantity of options, full information search and use rarely occurs (Simon, 1957; Bettman, Luce, and Payne, 1998).  Consequently there should exist any possibility of one stock (from \textit{early alphabet stock list}) giving better returns than other stocks in the sample (for example, \textit{later alphabet stocks}).

Through this research study, we attempt to examine several issues associated with the risk-return relationship in the context of penny stocks in the Indian stock market. Following are the major research questions that this paper is trying to answer:
\begin{enumerate}
	\item \textit{Does the effect of risk factors on expected returns vary across penny stocks with different characteristics?} We are interested in examining if penny stocks with higher risk attribute, such as high beta, exhibit return potential different from that of those penny stocks with lower risk attributes? If this holds true, it essentially explains the basic quantitative rule of risk-return relationship. And if \textit{vice versa} is true, the risk-return relationship in the context of penny stocks requires further investigation.
	\item \textit{If we see the examine the above problem from a risk-adjusted return perspective, can the results be any different?} Since risk-adjusted returns shall have less variation across risk factors, the observations should ideally change.
	Mainly we are considering \textit{firm size}, \textit{price-equity (P/E) ratio}, \textit{price-to-book value (P/B) ratio}, \textit{shareholding pattern}, \textit{beta}, and \textit{liquidity} as risk factors attributed to variation in returns. We are considering both raw and risk-adjusted returns in our case.
	\item To test the evidence the effect of alphabetism as one of the risk factors, we examine whether penny stocks with names starting with upper percentile of alphabets, such as A, B, C, ... and so on, exhibit different trading pattern in terms of trading volume compared to the stocks with names starting with lower percentile of alphabets, such as N, O, P, ... and so on?
\end{enumerate}

Although literature on financial economics and asset pricing have documented the role of several risk factors as well as behavioral factors in determining the return, very few studies focus on this issue in the context of the penny stocks. Moreover, this issue is rarely investigated in the Indian stock market. We believe that our article is one of the first attempts to look at how these factors interplay. Furthermore, the results are distinct in that they show how even for a small sample of listed penny stocks these traditional as well as unique risk factors are important, implying the behavioral tendency of such stocks.

\section{Measures and Methods}

\subsection{Data}

The data come from a cross-section of penny stocks listed in the National Stock Exchange of India. Our sample consists of a set of 167 listed penny stocks that are traded on the NSE electronic platform. These companies were selected to represent the penny stock universe as their share prices, as on $31^{st}$ July 2015, were below \rupee~10, which is the widely accepted cut-off price for classifying penny stocks in the Indian Stock Market. These stocks are then studied for a period of one year from August 2015 to July 2016.

The data variables for the study consist of annualized average returns (defined as $aar_{i} = 250  \Sigma^{N}_{d=1} r_{i,d}$), annualized average volatility (defined as $aav_{i} = 250  \Sigma^{N}_{d=1} r^{2}_{i,d}$), market capitalization, P/E Ratio, P/B Ratio, beta, liquidity and promoters' shareholding (defined as the percentage of shares held by promoters). The 365 day averages are computed over the period of one year from August 2015 to August 2016 and all the other data for the purpose are taken for July 2015. The cut-off yield on 364 Day Treasury Bills issued by the RBI has been used as the risk-free rate and the Sharpe Ratio, defined as $\frac{\text{365 day average return - risk-free rate}}{\text{365 day average volatility}}$  is used to represent the Risk Adjusted Returns. 

The data have been collected mainly from the CMIE-PROWESS database of the Center for Monitoring Indian Economy (CMIE) and the reports of the RBI available at the RBI Database on Indian Economy (RBI-DBIE).

\subsection{Methodology}

We first begin with testing for the effect of different microstructural and corporate governance related factors on the penny stock returns.

For our study, we consider one factor at a time. First, the stocks are ranked on the basis of the factor and then they are grouped them into two portfolios - top 50 percentile and the bottom 50 percentile.  We then run the statistical t-test on the annualized average returns ($aar_i$) and the annualized average risk-adjusted returns (365 Day Average Sharpe) for these two portfolios  to find out whether there is any significant difference between the returns of the portfolios formed. If a statistical significance is seen, then we conclude that the micro-structural factor considered has an effect in influencing the returns of these penny stocks. Further, we determine the nature of the influence, direct or inverse, by noting which portfolio gives the higher returns.

It is important to note here that we take the data for the factors on July 2015 while the returns and volatility data are taken for the period of August 2015 to July 2016. By doing so, we inherently are considering the annual average returns generated from following a buy and hold strategy where portfolios are formed in the month of July 2015 and held for one year.

Given the background of the research problems, we derive the research hypotheses and test with the data obtained as above. The baseline hypothesis is as following:
\begin{equation}
H_{0}: r_{ax} - r_{bx} = 0
\end{equation}
where, $r_{ax}$ is the mean (risk-adjusted) returns of the stocks belonging to the top 50 percentile after sorting in descending order w.r.t factor $x$, $r_{bx}$ is the mean (risk-adjusted) returns of the stocks belonging to the bottom 50 percentile after sorting in descending order w.r.t factor $x$, and $x$ includes risk factors such as firm size, P/E ratio, P/B ratio, price, shareholding pattern, beta, liquidity, and dummy for the alphabet with which the firm's name begins.

The statistical significance level (\textit{alpha}) is taken to be 0.05 and all the hypotheses are tested with two-tailed $t$-test carried out assuming unequal variances.

The functional alternate hypotheses ($H_{1}$) are stated as follows:

\begin{enumerate}[label=(\roman*)]
	\item The impact of firm size on average observed returns for top half of the sample firms are different than that of bottom half of the sample firms.
	\begin{subequations}

		\begin{equation}
		H_{1,1}: r_{\texttt{top 50}, \texttt{firm size}} - r_{\texttt{bottom 50}, \texttt{firm size}} \neq 0
		\end{equation}
	\item The impact of firm's P/E ratio on average observed returns for top half of the sample firms are different than that of bottom half of the sample firms.
		\begin{equation}
		H_{1,2}: r_{\texttt{top 50}, \texttt{P/E ratio}} - r_{\texttt{bottom 50}, \texttt{P/E ratio}} \neq 0
		\end{equation}
	\item The impact of firm's P/B ratio on average observed returns for top half of the sample firms are different than that of bottom half of the sample firms.
		\begin{equation}
		H_{1,3}: r_{\texttt{top 50}, \texttt{P/B ratio}} - r_{\texttt{bottom 50}, \texttt{P/B ratio}} \neq 0
		\end{equation}
	\item The impact of average promoters' shareholding on average observed returns for top half of the sample firms are different than that of bottom half of the sample firms.
		\begin{equation}
		H_{1,4}: r_{\texttt{top 50}, \texttt{shareholding}} - r_{\texttt{bottom 50}, \texttt{shareholding}} \neq 0
		\end{equation}
	\item The impact of average beta on average observed returns for top half of the sample firms are different than that of bottom half of the sample firms.
		\begin{equation}
		H_{1,5}: r_{\texttt{top 50}, \beta} - r_{\texttt{bottom 50}, \beta} \neq 0
		\end{equation}
	\item The impact of average liquidity on average observed returns for top half of the sample firms are different than that of bottom half of the sample firms.
		\begin{equation}
		H_{1,6}: r_{\texttt{top 50}, \texttt{liq}} - r_{\texttt{bottom 50}, \texttt{liq}} \neq 0
		\end{equation}
	\end{subequations}
\end{enumerate}

The hypotheses have been tested with both observed returns and risk-adjusted returns. The risk-adjusted returns here imply the returns adjusted with the Sharpe ratio.

The above-mentioned hypotheses intend to test:
\begin{subequations}
	\begin{equation}
H_{0}: \mu_{r_{ax}} = \mu_{r_{bx}}
\end{equation}
\begin{center}
	versus
\end{center}
\begin{equation}
H_{1}: \mu_{r_{ax}} \neq \mu_{r_{bx}}
\end{equation}
\end{subequations}
Our analysis uses the cross-sectional data observations, $x_{i,j}$ for stock $i$ and variable $j$, are independently and identically distributed random variables with normal distribution properties such that $N(\mu_{j}, \sigma^{2})$.

\section{Results and Discussions}

The traditional asset pricing framework as suggested by Sharpe (1964), Lintner (1965), and Black (1972) shows that the average returns and risk associated with any asset have strong relationship. The argument of the returns being a linear function of the associated risk have become as good as a power law. The relationship between large fluctuations in prices, trading volume and number of trades has also been established empirically (Gabrix et al., 2003). However, it has been observed that a typical risk-return relationship does not hold when it comes to penny stocks. This behavior may be attributed to several factors including information asymmetry, ownership structure, illiquidity, market microstructure issues, and so on. We basically examine whether stocks within a group of penny stocks exhibit particular characteristics. Some of the basic research issues that we are exploring are as following. Are the average returns on penny stocks with higher market capitalization (within the group) different from that on penny stocks with lower market cap? Do stocks with higher P/E ratio or P/B ratio show different average return generation capabilities compared to the stocks with lower P/E or P/B ratio? Does shareholding pattern make any difference in terms of return generation by a penny stock? Do those penny stocks with higher liquidity give any different returns than the penny stocks with less liquidity? These issues are investigated in the context of penny stocks listed in the Indian stock market.

We begin with presenting the descriptive statistics and statistical properties of our sample data. In Table \ref{fig:table1}, we show the calculated values of mean (along with standard error), standard deviation, maximum and minimum values, skewness and kurtosis, and the number of observations for each of the sample variables. The average return on our sample stock is about 6.27 percentage with a standard deviation of 0.27. As indicated by the range (maximum and minimum values) and skewness the data series is negatively skewed. Similarly, we report the central tendency and spread properties of our data variables, namely, firm size (as measured by market capitalization as on the date), P/E ratio, P/B ratio, shareholding pattern, beta, and liquidity. These values are obtained from the raw data collected from the sources as mentioned in the previous section.

Table \ref{fig:table2} shows the Pearson's correlations between pairs of sample variables. We can see that no variable is highly and significantly correlated with any other variable, hence there is no issue of multicollinearity. Firm size, P/E ratio, and P/B ratio are negatively correlated with the average return and the correlation coefficients are statistically significant. However, beta is positively correlated with average return at a 5 percent significance level. This intuitively supports the argument the positive association between risk and return on assets. Beta (as a measure of risk) is also positively correlated with firm size (at a conventional statistical significance level) which implies that firms with higher market cap carry high amount of risk (as suggested by traditional market theories). As expected, beta is negatively and significantly correlated with P/E ratio. However, we find no statistically significant correlation of liquidity with any of other sample variables. This is possibly the contemporaneous attribute of liquidity with other functions of market microstructure.

\begin{table}[h]
	\centering
	\caption{Summary statistics of the data variables}
	\begin{tabular}{lccccccc}
		\hline \hline
		\textit{Variable} & \textit{Mean} (SE) & \textit{Std. Dev.} & \textit{Max.} & \textit{Min.} & \textit{Skew.} (SE) & \textit{Kurt.} (SE) & N \\ \hline
		Avg. Ret. & 0.0627 (0.0209) & 0.2701 & 0.7500 & -1.120 & -0.700 (0.188) & 3.058 (0.374) & 167\\
		Firm Size & 488.92 (98.83) & 1273.33 & 9315.57 & 8.020 & 4.976 (0.188) & 27.446 (0.375) & 166\\
		P/E Ratio & 214.76 (79.18) & 592.52 & 3120.00 & 1.19 & 3.736 (0.319) & 13.961 (0.628) & 56\\
		P/B Ratio & 1.0835 (0.1866) & 2.1035 & 18.28 & 0.0400 & 5.321 (0.215) & 37.001 (0.427) & 127\\
		Shareholding & 42.5919 (1.6894) & 21.7665 & 93.60 & 0.100 & -0.230 (0.188) & -0.806 (0.375) & 166\\
		Beta & 0.8331 (0.0421) & 0.5441 & 3.670 & -0.530 & 0.922 (0.188) & 4.028 (0.374) & 167\\
		Liquidity & 148.25 (80.814) & 1041.22 & 12737.32 & 0.0100 & 11.064 (0.188) & 131.84 (0.375) & 166\\ \hline \hline
	\end{tabular}
	\label{fig:table1}
\end{table}

\begin{table}[h]
	\centering
	\caption{Pair-wise Pearson correlations among variables}
	\begin{tabular}{lccccccc}
		\hline \hline
		 & Avg. Ret. & Firm Size & P/E Ratio & P/B Ratio & Shareholding & Beta & Liquidity\\ \hline
		 Avg. Ret. & 1.000 &  &  &  &  &  & \\
		 Firm Size & -0.140* & 1.000 &   &  &  &   &  \\
		 P/E Ratio & -0.358*** & 0.332** & 1.000 &  &  &  & \\
		 P/B Ratio & -0.297*** & 0.259*** & 0.461*** & 1.000 &  &  & \\
		 Shareholding & 0.116 & 0.071 & 0.056 & 0.116 & 1.000 &  & \\
		 Beta & 0.172** & 0.309*** & -0.257* & -0.042 & -0.51 & 1.000 & \\
		 Liquidity & 0.071 & -0.047 & 0.088 & -0.033 & 0.059 & -0.054 & 1.000\\ \hline \hline
	\end{tabular}
	
	*, **, and *** indicate the significance (2-tailed) of correlation coefficients at 10\%, 5\%, and 1\%, respectively.
	\label{fig:table2}
\end{table}

With majority of sample data variables showing typical statistical and empirical properties, we further investigate whether our hypotheses as laid out before can be tested with the given data set.
Since the number of observations for our data variables are not same, we proceed with two-tailed $t$-test of the differences in means of data, to check whether two groups of data obtained from same sample/population and grouped on the basis of certain parameter(s) exhibit similar characteristic(s).

\begin{figure}[h]
	\centering
	\includegraphics[scale=0.8]{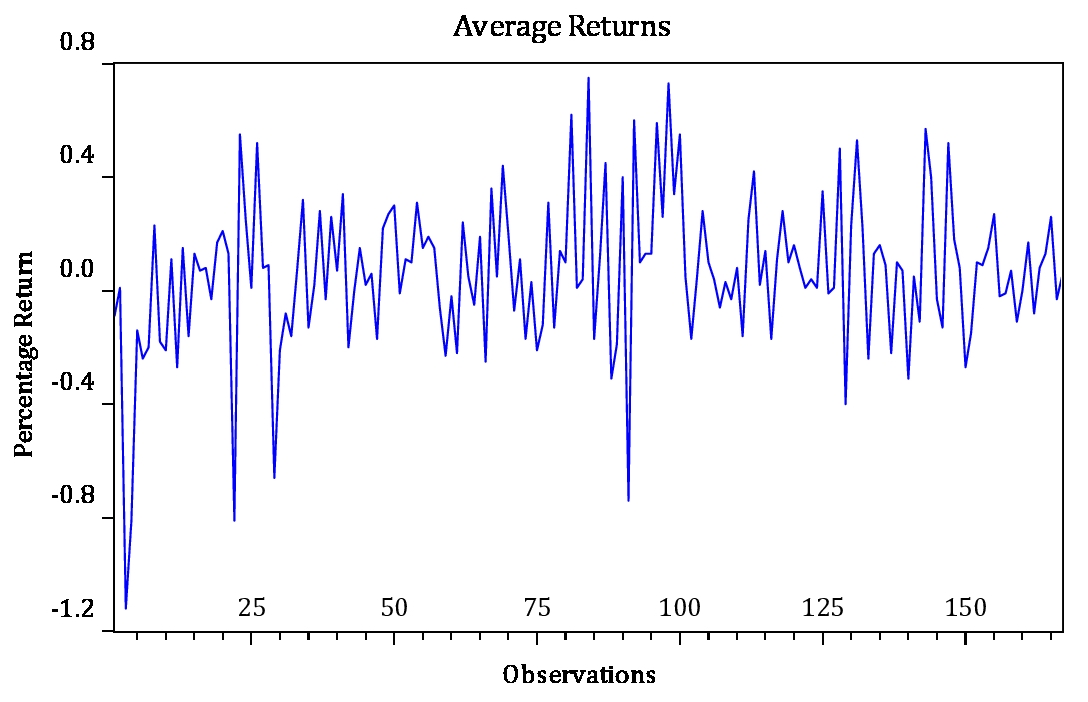} 
	\caption{The cross-sectional trend of average return on sample penny stocks}
	\label{fig:return1}
\end{figure}

Figure \ref{fig:return1} shows the sample-wide trend of cross-sectional return on selected penny stocks. The return data for a cross-section of penny stocks do not exhibit any anomalous trend as expected. Since the number of penny stocks listed and for which data are available in the context of the Indian stock market is very less, we can afford to have slightly dispersed data to be considered for our study. Kline (1998) proposed a rule of thumb for testing the normality of data values on the basis of which Hardigan et al. (2001) argue that any univariate skewness value greater than 3.0 and kurtosis higher than 10.0 may indicate the potential problem of normality in the data set. Data pertaining to firm size, P/E ratio, P/B ratio, and liquidity do not exhibit the normality assumption. We use logarithm of sample data for our relational analysis later. 

\begin{figure}[h]
	\centering
	\includegraphics[scale=0.9]{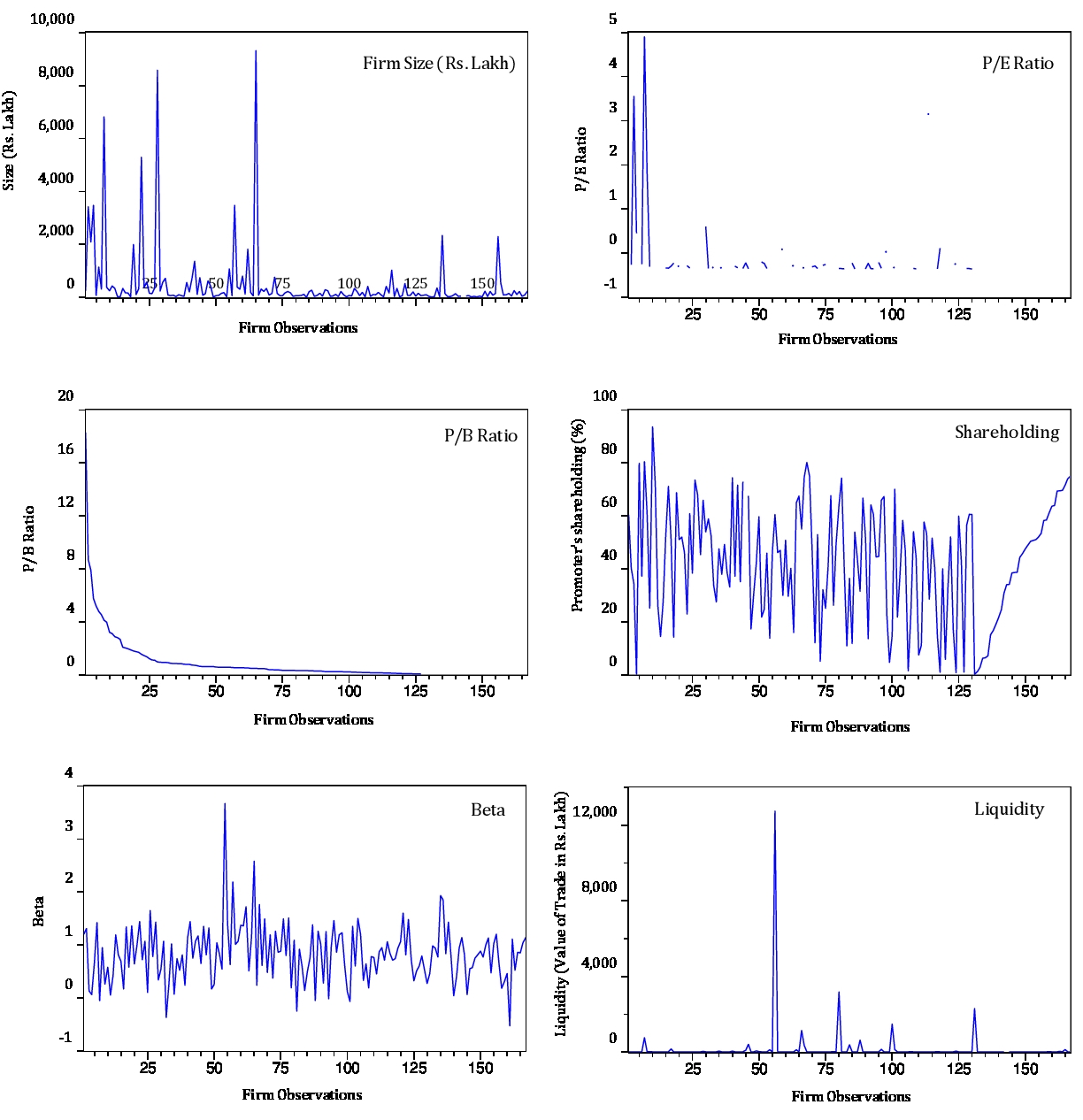} 
	\caption{The cross-sectional behavior of functions for our sample penny stocks}
	\label{fig:function1}
\end{figure}

We also show graphical representation of our other sample data variables, namely, firm size, P/E ratio, P/B ratio, shareholding pattern, beta, and liquidity. Figure \ref{fig:function1} shows the cross-sectional behavior of data values. As expected, the data exhibits wide variances across the sample variables. However, this should not deter us to use these data for testing our hypotheses. We argue that since our primary objective is to see whether top 50 percent and bottom 50 percent of stocks exhibit different behavior across variable, \textit{independently}, we can well use these data for our analysis. As of now, we don't try to show any causal or directional relationship as such, for example through regression, hence these data sets are appropriate for testing our hypotheses. We further aim to use a section of the data set for more sophisticated modeling to show functional relationship of expected average return with other functions that would be significantly driving the return properties of the sample penny stocks.

Now we shall discuss the results obtained from the statistical tests employed on our sample data. Table \ref{fig:table3} reports $t$-test statistics obtained using the sample data for all the variables. Our statistical test results show that firm size, P/E ratio, and P/B ratio are statistically significant with $t$-statistics of -3.1518 ($p$-value: 0.001), -2.4303 ($p$-value: 0.019), and -2.2264 ($p$-value: 0.027), respectively. Other variables such as shareholding ($t$-stat: 1.1963; $p$-value: 0.2334), beta ($t$-stat: 1.2401; $p$-value: 0.2168), and liquidity ($t$-stat: -0.7202; $p$-value: 0.472) are statistically insignificant.

\begin{table}[h]
	\centering
	\caption{$t$-test statistics for hypotheses testing: Observed returns}
	\begin{tabular}{lcccccc}
		\hline \hline
		\textit{Hypothesis} & $t$-stat. & \textit{df} & $p$-value & \textit{Mean Difference} & Lower CI (95\%) & Upper CI (95\%)\\ \hline
		$H_{1,1}$: Firm Size & -3.1518 & 158 & 0.001 & 0.1277 & 0.197 & 0.723 \\
		$H_{1,2}$: P/E Ratio & -2.4303 & 45 & 0.019 & 0.2039 & 0.392 & 0.835 \\
		$H_{1,3}$: P/B Ratio & -2.2264 & 124 & 0.027 & 0.1100 & 0.194 & 0.503 \\
		$H_{1,4}$: Shareholding & 1.1963 & 155 & 0.2334 & 0.0419 & 0.327 & 0.784 \\
		$H_{1,5}$: Beta & 1.2401 & 156 & 0.2168 & 0.0124 & 0.269 & 0.486 \\
		$H_{1,6}$: Liquidity & -0.7202 & 150 & 0.472 & -0.0211 & -0.072 & 0.416 \\ \hline \hline 
	\end{tabular}
	\label{fig:table3}
\end{table}

\textit{Firm size and return on penny stocks}: The significance of $t$-test results suggests that within a sample of penny stocks listed in the Indian stock market, there exists a negative association between firm size and average returns (as indicated by the correlation also). We report that the top 50 percent of penny stocks, sorted on the basis of size as denoted by market capitalization, have generated lower returns compared to the stocks from the bottom 50 percent. It implies that smaller is the firm size, higher is the average returns. This seems very much aligned with the classical theory that suggests smaller firms have more information asymmetry and carry higher risk, hence the expected return must be higher when compared to bigger firms. The empirical model such as Fama-French three factor model also states the same. Our simple comparison of average returns support the hypothesis.

uWe show that beta is positively correlated with the firm size (market cap) which is counter-intuitive as it points to the inference that larger firms are riskier. Beta here does not measure the fundamental risk which is higher for the smaller firms but only sees volatility and thus our paper points to the failure of beta to measure risk fully. The size effect is not due to the so called risk measured by beta. There is some other cause on why lower market cap penny stocks earn higher returns. It appears that the fundamental risk or asymmetry not fully captured by the beta. Essentially then, a portfolio of low market cap penny stocks have higher returns as well as low volatility which is great from investors point of view. Also, this reverse trend can explain why beta is not statistically significant in $t$-test despite being correlated with returns.

\textit{P/E ratio and P/B ratio}: A statistically significant $t$-statistic suggest that the returns on stocks belonging to bottom 50 percent when sorted on their P/E ratios are substantially higher than the returns on penny stocks from the top 50 percent of the P/E ratio-sorted stock group. A similar behavior is observed when we sorted the stocks on the basis of P/B ratio and test for the differences in means of returns on top 50 percent stocks and bottom 50 percent stocks. These two functions, namely, P/E and P/B ratios are established risk factors validated both theoretically and empirically across the markets and our results support the hypotheses laid out earlier.

The $t$-test results obtained for shareholding pattern, beta, and liquidity do not yield favorable results, suggesting that the returns on top 50 percent penny stocks are statistically and significantly indifferent from the returns on bottom 50 percent penny stocks when sorted on the basis or these functions. It appears that since most of the penny stocks belong to the firms held largely by the families and small business operators, shareholding pattern turns out to be an insignificant differentiator of returns on such stocks. Similarly, stocks when sorted on the basis of beta do not exhibit different pattern across top and bottom 50 percentiles. Liquidity factor is another redundant input as the sample penny stocks are very illiquid and rarely traded. It comes as no surprise to us as liquidity can influence returns only when it is heavily traded, resulting into high trading volume and price fluctuations.

\begin{table}[h]
	\centering
	\caption{$t$-test statistics for hypotheses testing: Risk-adjusted returns}
	\begin{tabular}{lcccccc}
		\hline \hline
		\textit{Hypothesis} & $t$-stat. & \textit{df} & $p$-value & \textit{Mean Difference} & Lower CI (95\%) & Upper CI (95\%)\\ \hline
		$H_{2,1}$: Firm Size & -3.3024 & 158 & 0.001 & 0.0442 & 0.189 & 0.716 \\
		$H_{2,2}$: P/E Ratio & -2.4263 & 45 & 0.021 & 0.0771 & 0.396 & 0.794 \\
		$H_{2,3}$: P/B Ratio & -2.1110 & 124 & 0.037 & 0.0353 & 0.179 & 0.494 \\
		$H_{2,4}$: Shareholding & 1.1934 & 155 & 0.235 & 0.1038 & 0.308 & 0.780 \\
		$H_{2,5}$: Beta & -0.7202 & 156 & 0.472 & 0.0163 & 0.263 & 0.461 \\
		$H_{2,6}$: Liquidity & -1.954 & 150 & 0.277 & 0.0912 & -0.106 & 0.409 \\ \hline \hline 
	\end{tabular}
	\label{fig:table4}
\end{table}

In Table \ref{fig:table4}, we show the results of $t$-tests carried out with a slightly modified measure of returns. We adjusted returns on penny stocks with Sharpe ratio, and use this risk-adjusted return to validate our argument whether risk-adjusted returns show any different behavior across our sample stocks. The results are consistent and only firm size, P/E and P/B ratios are statistically significant differentiators. Other functions such as shareholding, beta, and liquidity are insignificant. 

\textit{Alphabetism, behavioral biases and trading volume}: We further test our hypothesis that stocks with names starting with early/top 50 percent of alphabets (such as A, B, C, ..., L, M) have shown better average returns compared to stocks with later (bottom 50 percent) alphabets (such as N, O, ..., Y, Z). The theoretical justification of this hypothesis comes from early work by satisficing and the status-quo bias. Empirical and experimental research in psychology and economics shows that when provided with a choice between a significantly large number of alternatives, each with different risk and return characteristics, any economic agent tends to satisfice, whereby search ceases after an “economically acceptable” option is found, even if a better option could be found through continued search (Simon, 1956; Payne, 1976; Caplin, Dean, and Martin, 2011). Thus, in financial markets, when stock traders browse through the lists of stocks possibly available for trading, they are more likely to take positions against/for (either buy or sell) stocks appearing toward the beginning of the list, indicating early alphabets when arrange alphabetically. Therefore, initial ordering and presentation of alternatives significantly affects which stocks are selected for purchase or sale (Itzkowitz et al, 2015).

However, out simple two-tailed $t$-test statistic ($t$-stat: 0.3092; $p-$value: 0.7576) rejects the hypothesis of the presence of alphabetism in our sample. One reason could be a small set of stocks being considered in our study and the associated lack of data. It is hypothesized that investors find is easy to look through small set of stocks available for investing and, therefore, no alphabetism-induced bias can be established. However, such an exercise requires further investigation with larger sample and datasets.

\section{Concluding Remarks}

It is observed that Firm Size (measured in terms of market capitalization), P/E ratio, and P/B ratio significantly influence the returns with penny stocks having lower values of these three factors generating higher raw as well as risk adjusted returns. Penny stocks with lower prices generate higher raw returns but the premium vanishes after adjusting for risk while promoters' shareholding patterns, risk attributes (measured by beta) and liquidity do not show statistically significant influence on returns of the penny stocks.

This points to the existence of the much debated size effect, price-to-earnings ratio effect and the book-to-market effect in the Indian penny stock universe. It is possible to generate profits, even after adjusting for risk, from these stocks in the Indian stock market by forming portfolios on the basis of these factors. Though this paper explores a buy-and-hold strategy extending for a year, there can be other trading strategies which can utilize the existence of the size and effect of P/E and P/B ratios to generate adequate profits. 

\newpage

\section*{References}
\begin{enumerate}
	\item  Gabrix, X., P. Gopikrishnan, V. Plerou, and H. E. Stanley (2003). A theory of power-law distributions in financial market fluctuations. \textit{Nature}, Vol. 423, pp. 267-270.
	\item Itzkowitz, J. J. Itzkowitz, and S. Rothbort (2015). ABCs of Trading: Behavioral Biases affect Stock Turnover and Value. \textit{Review of Finance}, doi: 10.1093/rof/rfv012.
	\item Liu, Q., S. G. Rhee, and L. Zhang (2011). On the Trading Profitability of Penny Stocks. \textit{24th Australasian Finance and Banking Conference 2011 Paper}. Available at SSRN (http://ssrn.com/abstract=1917300).
	\item Moor, L. D., and P. Sercu (2015). The smallest stocks are not just smaller: Global evidence. \textit{The European Journal of Finance}, Vol. 21, No. 1, pp. 51-70.
	\item Murdock, B. B. (1962). The serial position effect of free recall. \textit{Journal of Experimental Psychology}, Vol. 64, pp. 482–488.
	\item Nofsinger, J. F., and A. Varma (2014). Pound wise and penny foolish? OTC stock investor behavior. \textit{Review of Behavioral Finance}, Vol. 6 Iss. 1, pp. 2-25.
	\item Payne, J. W. (1976). Task complexity and contingent processing in decision making: an information	search and protocol analysis. \textit{Organizational Behavior and Human Performance}, Vol. 16, pp. 366–387.
	\item Simon, H. A. (1956). Rational choice and structure of the environment. \textit{Psychological Review}, Vol. 63, pp.	129–138.
	\item Simon, H. A. (1957). \textit{Models of Man: Social and Rational}, John Wiley and Sons, Inc., New York,	New York.
	\item Subrahmanyam, A. (2010). The Cross-Section of Expected Stock Returns: What Have We Learnt from the Past Twenty-Five Years of Research? \textit{European Financial Management}, Vol. 16, No. 1, pp. 27-42.
	\item Tversky, A. and Kahneman, D. (1973). Availability: a heuristic for judging frequency and probability. \textit{Cognitive Psychology}, Vol. 5, pp. 207–232.
\end{enumerate}
\end{document}